\documentclass[journal,comsoc]{IEEEtran}
\usepackage[T1]{fontenc}
\usepackage{bm}
\usepackage{subfig}
\usepackage{amsmath}
\usepackage{stfloats}
\usepackage{graphicx}
\usepackage{cite}
\newcommand{\RNum}[1]{\uppercase\expandafter{\romannumeral #1\relax}}
\ifCLASSINFOpdf
\else
\fi

\begin{document}

\title{A Novel Demodulation and Estimation Algorithm for Blackout Communication: Extract Principal Components with Deep Learning }

\author{Haoyan~Liu,
        Yanming~Liu,
        Min~Yang
	  and Xiaoping~Li
\thanks{The authors are the School of Aerospace Science and Technology, Xidian University, Xi’an 710071, China}
}


\maketitle

\begin{abstract}
For reentry or near space communication, owing to the influence of the time-varying plasma sheath channel environment, the received IQ baseband signals are severely rotated on the constellation. Researches have shown that the frequency of electron density varies from 20kHz to 100 kHz which is on the same order as the symbol rate of most TT\&C communication systems and a mass of bandwidth will be consumed to track the time-varying channel with traditional estimation. In this paper, motivated by principal curve analysis, we propose a deep learning (DL) algorithm which called symmetric manifold network (SMN) to extract the curves on the constellation and classify the signals based on the curves. The key advantage is that SMN can achieve joint optimization of demodulation and channel estimation. From our experiment results, the new algorithm significantly reduces the symbol error rate (SER) compared to existing algorithms and enables accurate estimation of fading with extremely high bandwith utilization rate.

\end{abstract}


\IEEEpeerreviewmaketitle

\section{Introduction}

\IEEEPARstart{W}{hen} the vehicles fly in the atmosphere at supersonic speed, the air surrounding the vehicles is dissociated and ionized by tremendous heat. The ionized gas, called the plasma sheath, contains massive free electrons that absorb, reflects or scatters electromagnetic (EM) waves. Yang analyze the amplitude fading and phase shift of the IQ signals caused by time-varying plasma, while they find that the received signals will be rotated along curves on constellation \cite{10.1063/1.4950694}.

Recently, some researches have applied DL in channel estimation\cite{8353153}\cite{8403666}. T. J. O'Shea, who has made many contributions to combine DL and communication signal processing, points out that the potential of DL in the physical layer mainly comes from the following two aspects\cite{8054694}: First, most communication signal processing methods are under the assumption of the existing theoretical channel model (e.g., liner, stationary or Gaussian and so on). However, wireless channels are complicated in practice especially for plasma sheath. In this condition, DNN can provide better performance to fit authentic wireless channels. Secondly, the entire signal processing chain is modularized and each module is optimized independently. This is a greedy approach which could not guarantee end-to-end optimization of the system with the evidence that the joint code modulation can achieve more gain than modulation after encoding\cite{1056454}. 

In this paper, our work primarily focuses on demodulation and fading estimation in time-varying plasma. The fact that the received baseband signals lie on multiple curves motivates us to extract the principal components (i.e., fading curves)\cite{10.1080/01621459.1989.10478797} and then determine which curve each signal belongs to. We design a novel DNN structure called symmetric manifold network (SMN) trained with semi-supervised learning. Our basic idea is to couple estimation and demodulation by extracting the fading curve with the existing soft decision information and updating the soft decision information with the new fading curve iteratively. We prove that this is an approximate maximum likelihood estimation for the received signals. Experiments show that our algorithm can obtain lower symbol error rate than traditional maximum likelihood receivers and some supervised learning algorithms by fewer training sequences. Simultaneously, the results show that the channel estimation method is robust to noise.

\section{Time-varying Plasma Sheath Channel}
The plasma sheath can be considered a dispersive and lossy medium. The radio wave propagation characteristics are related to the carrier frequency $\omega$, plasma frequency $\omega_{p}$, and collision frequency $v_{e n}$. Affected by the time-varying electron density $n_{e}(t)$, the plasma frequency $\omega_{p}(t)$ can be expressed as\cite{10.1063/1.4907904} 
\begin{equation}
	\omega_{p}(t)=\sqrt{\frac{n_{e}(t) e^{2}}{\varepsilon_{0} m_{e}}}
	\label{eq1}
\end{equation} 
where $e$ is the electron charge, $\varepsilon_{0}$  is the dielectric constant in vacuum, and $m_{e}$ is the electron mass. The  complex dielectric coefficient $\varepsilon_{r}(t)$ is given by 
\begin{equation}
	\varepsilon_{r}(t)=1-\frac{\omega_{p}^{2}(t)}{\omega^{2}+v_{e n}^{2}}-j 
	\frac{v_{e n}^{2}}{\omega} \frac{\omega_{p}^{2}(t)}{\omega^{2}+v_{e n}^{2}}
	\label{eq2}
\end{equation}
the propagation vector can be expressed as 
\begin{equation}
	k(t)=\frac{\omega}{c} \sqrt{\varepsilon_{r}(t)}=\beta(t)-j \alpha(t)
	\label{eq3}
\end{equation} 
with (\ref{eq2}), (\ref{eq3}), the attenuation coefficient $\alpha(t)$ and phase-shift coefficient $\beta(t)$ can be derived as 
\begin{equation}
	\begin{aligned}
		\alpha(t)&=\frac{\omega}{\sqrt{2} c} \operatorname{sqrt}\left(\frac{\omega_{p}^{2}(t)}{\omega^{2}+v_{en}^{2}}-1\right. \\
		&\left.+\sqrt{\left(1-\frac{\omega_{p}^{2}(t)}{\omega^{2}+v_{c n}^{2}}\right)^{2}+\left(\frac{v_{e n}^{2}}{\omega} \frac{\omega_{p}^{2}(t)}{\omega^{2}+v_{e n}^{2}}\right)^{2}}\right)
		\label{eq4}
	\end{aligned}
\end{equation}
\begin{equation}
	\begin{aligned}
		\beta(t)&=\frac{\omega}{\sqrt{2} c} \operatorname{sqrt}\left(1-\frac{\omega_{p}^{2}(t)}{\omega^{2}+v_{e n}^{2}}+\right. \\
		&\left.\sqrt{\left(1-\frac{\omega_{p}^{2}(t)}{\omega^{2}+v_{e n}^{2}}\right)^{2}+\left(\frac{v_{e n}^{2}}{\omega} \frac{\omega_{p}^{2}(t)}{\omega^{2}+v_{e n}^{2}}\right)^{2}}\right)
		\label{eq5}
	\end{aligned}
\end{equation}

\begin{figure}[!t]
	\centering
	\subfloat[]{
		\label{fig6_a}
		\begin{minipage}[t]{0.5\linewidth}
			\centering
			\includegraphics[width=2.0in]{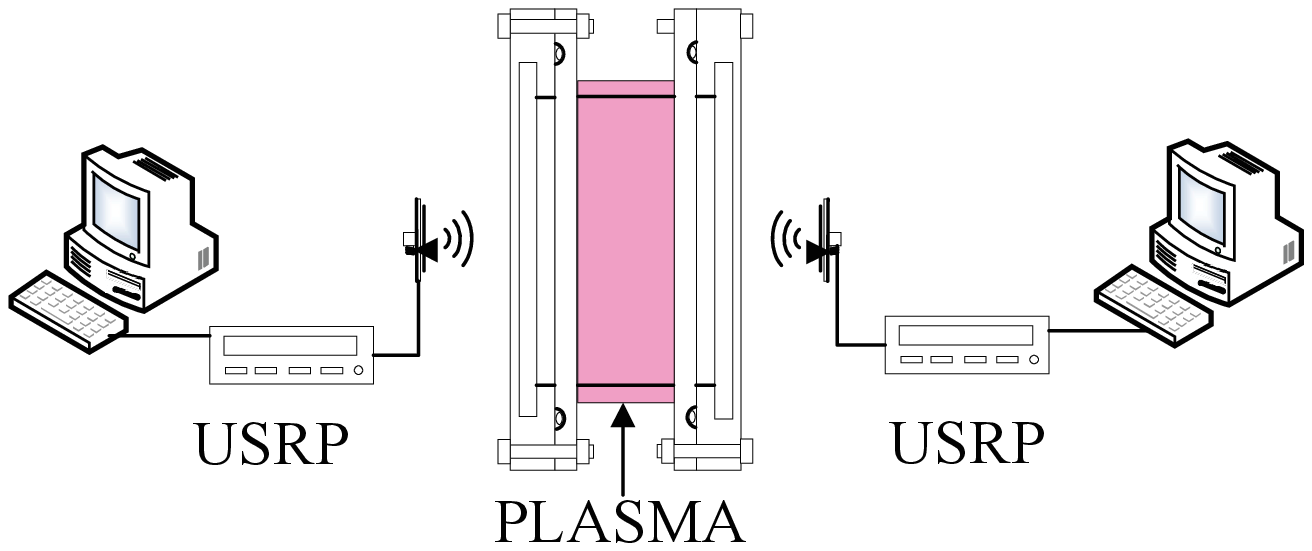}
		\end{minipage}
	}
	\subfloat[]{
		\label{fig6_b}
		\begin{minipage}[t]{0.5\linewidth}
			\centering
			\includegraphics[width=1in]{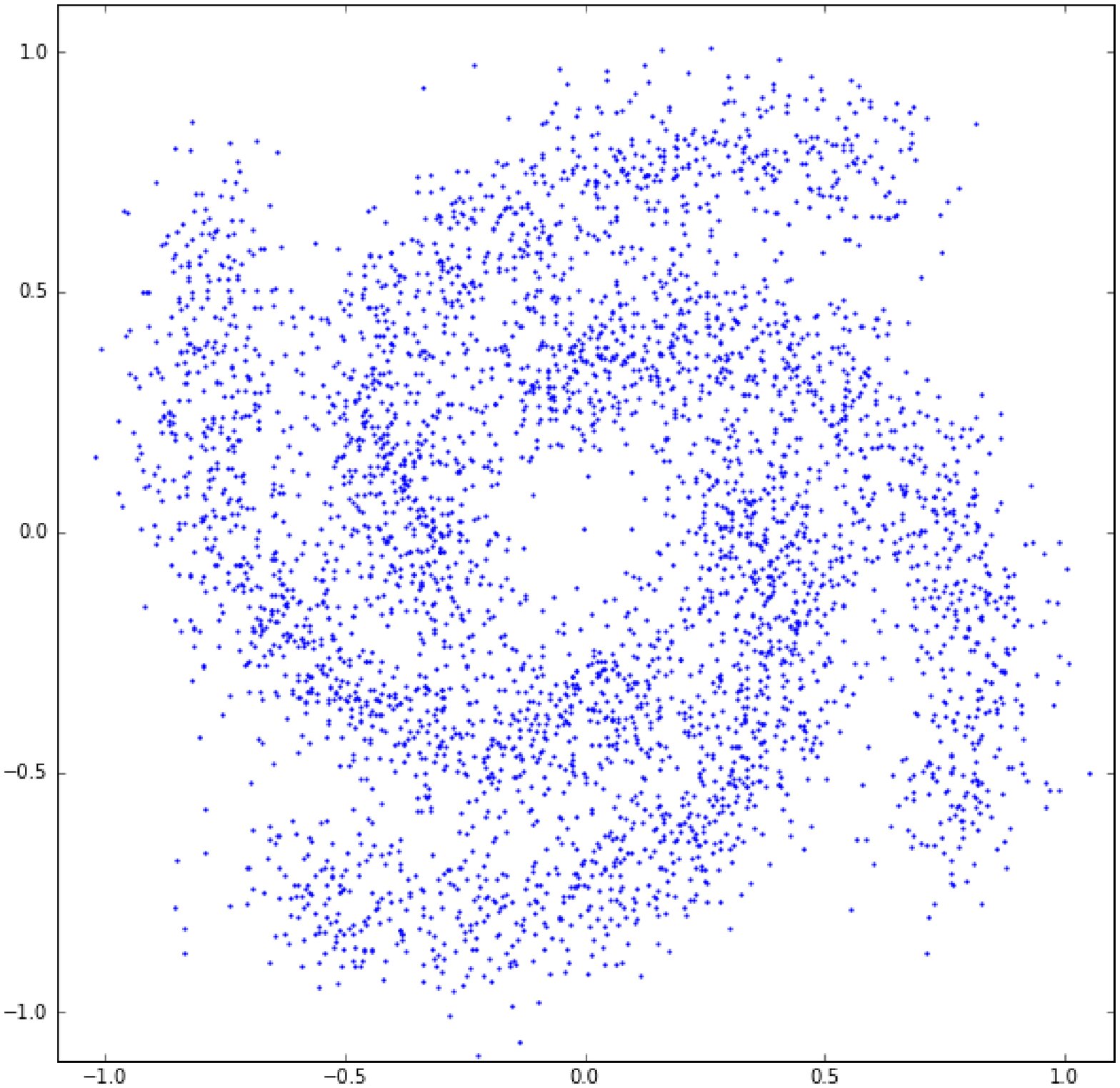}
		\end{minipage}
	}
	\caption{(a) Physical simulation system; (b) The received signals.}
	\label{fig6}
\end{figure}

For M-ary IQ modulation system, the received baseband signal can be expressed as
\begin{equation}
	\begin{aligned}
	 \bm{y}_{i} &=e^{-j k\left(n_{e}\right) z} \bm{x}_{i}+\bm{n}_{i}\\ 			&=e^{-\alpha\left(n_{s}\right) z} e^{-j \beta\left(n_{s}\right) z} \bm{x}_{i}+\bm{n}_{i} \\ &=\bm{s}\left(n_{e}\right) \bm{x}_{i}+\bm{n}_{i}
	 \label{eq6}
	\end{aligned}
\end{equation}
where $\bm{x}_{i} \in\left[\bm{x}_{1}, \bm{x}_{2}, \cdots, \bm{x}_{K}\right], K=2^{M}$ is the transmitted complex baseband signal, $\bm{y}_{i}$ is the received signal, and $\bm{n}_{i} \sim \mathcal{N}\left(0, \sigma_{n}^{2}\right)$ is complex Gaussian white noise. $h=e^{j t}$ is the channel fading, $z$ is the thickness of the plasma sheath. It can be seen that $n_{e}(t)$ causes the time-varying amplitude fading and phase shift of $\bm{y}_{i}$ . As shown in Fig.\ref{fig6}, we simulate the effect of time-varying plasma on QPSK signals by universal software radio peripheral (USRP) and plasma generator. That $\bm{y}_{i}$ will be Gaussian along the curve set $S=\left[s_{1}, s_{2}, \cdots, s_{K}\right]$ on the constellation motivates us to demodulate by extracting the fading curve.
\section{Technical Approach}

\subsection{Principal Curve}
Principal curve is a generalization of the first linear component in nonlinear condition. The curve in high dimensional space is regarded as one dimensional manifold embedded in Euclidean space. Consequently, it can be described as with a single variable $\lambda$ and a coordinate function which is denoted as $\bm{f}(\lambda)=\left(f_{1}(\lambda), f_{2}(\lambda), \cdots, f_{d}(\lambda)\right)$. $\bm{f}$ is by definition a smooth curve if the coordinate functions $\left[f_{1}, f_{2}, \cdots, f_{d}\right]$ are smooth\cite{10.1080/01621459.1989.10478797}.

\begin{figure}[!t]
	\centering
	\subfloat[]{
		\label{fig1_a}
		\begin{minipage}[t]{0.5\linewidth}
			\centering
			\includegraphics[width=1.6in]{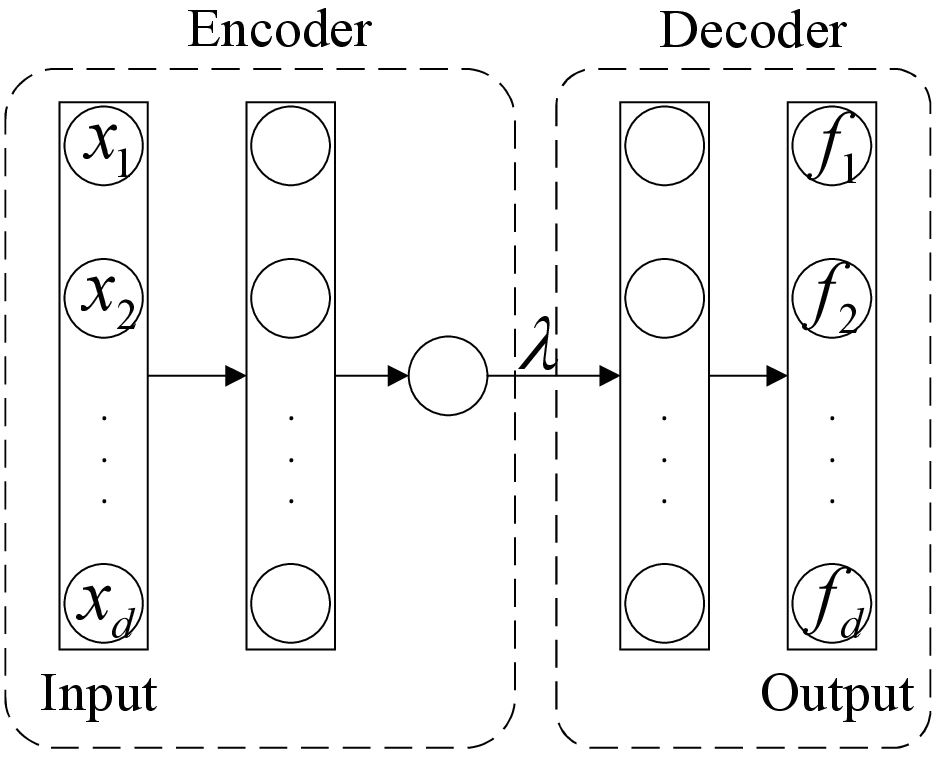}
		\end{minipage}
	}
	\subfloat[]{
		\label{fig1_b}
		\begin{minipage}[t]{0.4\linewidth}
			\centering
			\includegraphics[width=1.3in]{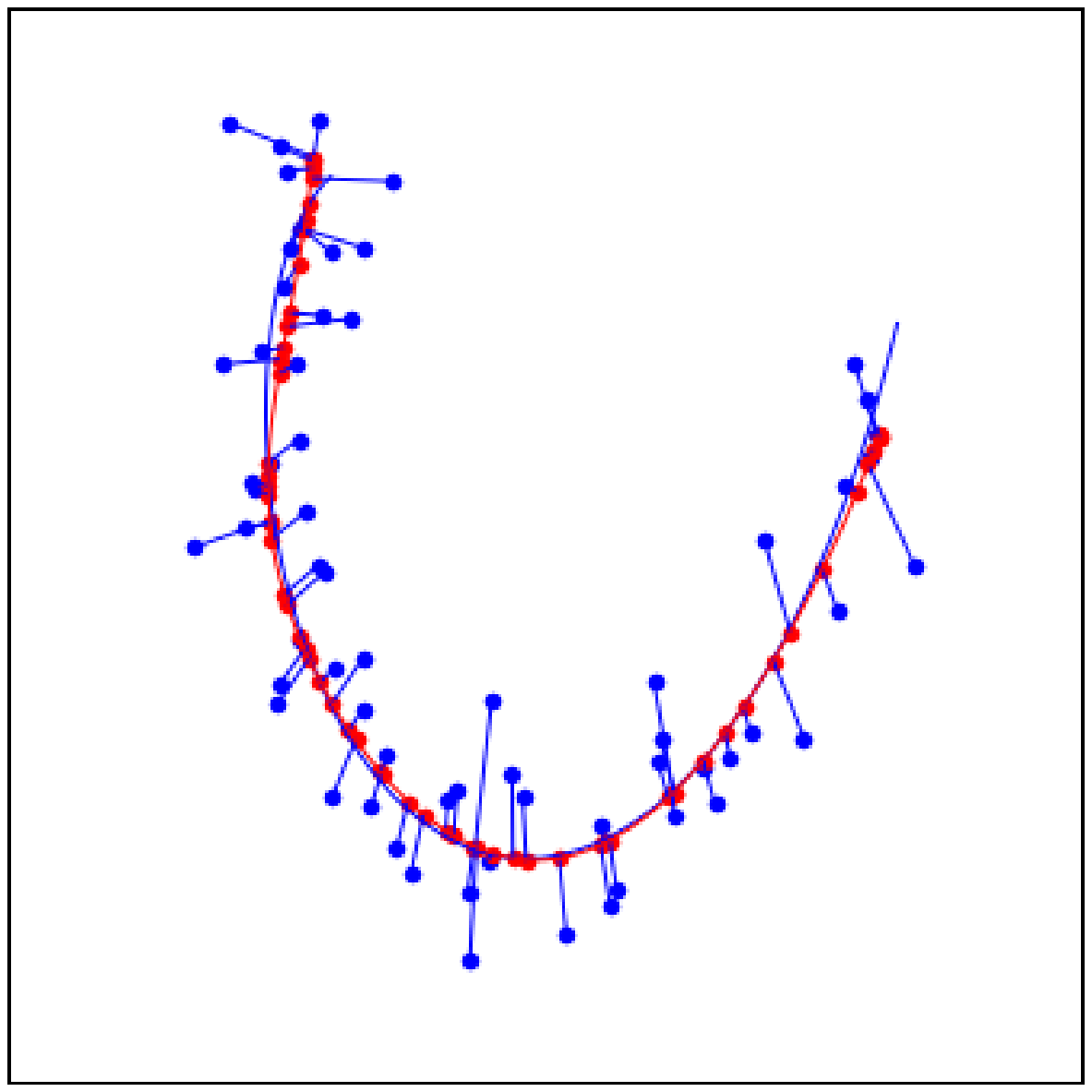}
		\end{minipage}
	}
	\caption{(a) Structure of autoencoder; (b) Simulation results for the principal curve. The blue dots, the red dots, the blue curve and the red curve represent the noisy data, the projection points, the generative curve and the learned curve respectively.}
	\label{fig1}
\end{figure}

Based on the principle of universal approximation, our intention is to construct a parametric model of the curve by DNN. As shown in Fig.\ref{fig1_a}, a five-layer undercomplete autoencoder\cite{Bengio:2006:GLT:2976456.2976476} is implemented to fit the curve. The encoder is used to compute $\lambda$, and the decoder is used to approximate $\bm{f}$. Fig.\ref{fig1_b} shows the projection points of the input on the curve can be obtained by minimizing the mean square error.

\subsection{Approximate Maximum Likelihood Estimation }

Donate that $\bm{X}=\left[\boldsymbol{x}_{1}, \bm{x}_{2}, \cdots, \bm{x}_{m}\right]$ is the transmitted sequence and $\bm{Y}=\left[\bm{y}_{1}, \bm{y}_{2}, \cdots, \bm{y}_{m}\right]$ is the received sequence. The log-likelihood function of $\bm{Y}$ can be decomposed as
\begin{equation}
	\begin{aligned} 
		\ln &p(\boldsymbol{Y})=\ln \sum_{\boldsymbol{X}} p(\boldsymbol{Y}, \boldsymbol{X}) \\ 
		&=\sum_{\boldsymbol{X}} q(\boldsymbol{X}) \ln \left[\frac{p(\boldsymbol{Y} | \boldsymbol{X}) p(\boldsymbol{X})}{q(\boldsymbol{X})}\right]- \sum_{\boldsymbol{X}} q(\boldsymbol{X}) \ln \left[\frac{p(\boldsymbol{X} | \boldsymbol{Y})}{q(\boldsymbol{X})}\right] \\ &=\mathcal{L}(\boldsymbol{X}, \boldsymbol{Y})+D_{K L}[q(\boldsymbol{X}) \| p(\boldsymbol{X} | \boldsymbol{Y})] 
		\label{eq7}
	\end{aligned}
\end{equation}
where $q(\bm{X})$ is an arbitrary distribution of $\bm{X}$, $D_{K L}$ denotes the Kullback Leibler distance. The fact that $D_{K L} \geq 0$ indicates that $\mathcal{L}(\boldsymbol{X}, \boldsymbol{Y})$ is the lower bound of  $\ln p(\boldsymbol{Y})$. If and only if $q(\boldsymbol{X})=p(\boldsymbol{X} | \boldsymbol{Y})$, $\ln p(\boldsymbol{Y})$ takes the maximum value. Consequently, an effective method is to maximizing $\mathcal{L}(\boldsymbol{X}, \boldsymbol{Y})$ instead of $\ln p(\boldsymbol{Y})$.

In (\ref{eq7}), the $p(\boldsymbol{Y} | \boldsymbol{X})$ could be problematic due to the time-varying $\bm{h}$
\begin{equation}
	\begin{aligned}
	p(\boldsymbol{Y} | \boldsymbol{X})&=\prod_{i}^{m} p\left(\boldsymbol{y}_{i} | \boldsymbol{x}_{i}\right) 
	=\prod_{i}^{m} \frac{1}{\pi \sigma_{n}^{2}} \exp \left(\frac{-\left\|\boldsymbol{y}_{i}-\boldsymbol{h} \boldsymbol{x}_{i}\right\|_{2}^{2}}{\sigma_{n}^{2}}\right)
	\label{eq8}
	\end{aligned}
\end{equation}
where $\|\cdot\|_{2}^{2}$ denotes the Euclidean distance. From the perspective that $p\left(\boldsymbol{y}_{i} | \boldsymbol{x}_{i}\right)$ can be considered as the probability of $\bm{y}_i$ given the curve $\bm{s}_k$, the $p\left(\boldsymbol{y}_{i} | \boldsymbol{x}_{i}\right)$ is rewritten as
\begin{equation}
	p\left(\boldsymbol{y}_{i} | \boldsymbol{s}_{k} ; \boldsymbol{\theta}\right)=\frac{1}{\pi \sigma_{n}^{2}} \exp \left(\frac{-\left\|\boldsymbol{y}_{i}-\lambda_{f_{k}}\left(\boldsymbol{y}_{i} ; \boldsymbol{\theta}\right)\right\|_{2}^{2}}{\sigma_{n}^{2}}\right)
	\label{eq9}
\end{equation}
where $\lambda_{f_{t}}\left(\boldsymbol{y}_{i}\right)$ is the projection coordinate of $\bm{y}_i$ on $\bm{s}_k$ and $\boldsymbol{\theta}$ is the parameter of $\lambda_{f_{t}}\left(\boldsymbol{y}_{i}\right)$. Since $\lambda_{f_{t}}\left(\boldsymbol{y}_{i}\right)$ is infeasible in practice, it can be implemented by DNN.

In communication systems, it is generally considered that the prior probabilities $p(\bm{X})$ are equal. The $\mathcal{L}(\boldsymbol{X}, \boldsymbol{Y})$ can be optimized with expectation maximization (EM) algorithma\cite{Dempster77maximumlikelihood} as follows \\
E-step: Update $q(\bm{X})$  which can be regard as posterior probability matrix $\bm{W}$ with the current $\boldsymbol{\theta}$ and $\sigma_{n}^{2}$
\begin{equation}
	\begin{aligned}
	q\left(\boldsymbol{x}_{i k}\right)=\boldsymbol{W}_{i k}&=p\left(\boldsymbol{s}=k | y_{i} ; \boldsymbol{\theta}, \sigma_{n}^{2}\right)\\ 
	&=\frac{p\left(\boldsymbol{s}=k | y_{i} ; \boldsymbol{\theta}, \sigma_{n}^{2}\right) p(\boldsymbol{s}=k)}{\sum_{l=1}^{K} p\left(\boldsymbol{s}=l | y_{i} ; \boldsymbol{\theta}, \sigma_{n}^{2}\right) p(\boldsymbol{s}=l)} \\&=\frac{\exp \left(\frac{-\left\|\boldsymbol{y}_{i}-\lambda_{f_{k}}\left(\boldsymbol{y}_{i} ; \boldsymbol{\theta}\right)\right\|_{2}^{2}}{\sigma_{n}^{2}}\right)}{\sum_{l=1}^{K} \exp \left(\frac{-\left\|\boldsymbol{y}_{i}-\lambda_{f_{k}}\left(\boldsymbol{y}_{i} ; \boldsymbol{\theta}\right)\right\|_{2}^{2}}{\sigma_{n}^{2}}\right)}
	\label{eq10}
	\end{aligned}
\end{equation}\\
M-step: Re-estimate $\boldsymbol{\theta}$ and $\sigma_{n}^{2}$
\begin{equation}
	\begin{aligned} 
		\boldsymbol{\theta}, \sigma_{n}^{2} &=\underset{\theta, \sigma_{n}^{2}}{\arg \max } \mathcal{L}\left(\boldsymbol{X}, \boldsymbol{Y} ; \boldsymbol{\theta}, \sigma_{n}^{2}\right) \\ &=\underset{\theta, \sigma_{n}^{2}}{\arg \max } \sum_{i} \sum_{k} \bm{W}_{i k} \log \frac{p\left(y_{i} | \boldsymbol{s}=k ; \boldsymbol{\theta}, \sigma_{n}^{2}\right) p(\boldsymbol{s}=k)}{W_{i k}} \\ &=\underset{\theta, \sigma_{n}^{2}}{\arg \max } \sum_{i} \sum_{k} \bm{W}_{i k} \log p\left(y_{i} | \boldsymbol{s}=k ; \boldsymbol{\theta}, \sigma_{n}^{2}\right) 
		\label{eq11}
	\end{aligned}
\end{equation}\\ 
(\ref{eq11}) can be further derived as 
\begin{equation}
	\begin{aligned}
		\boldsymbol{\theta}&=\underset{\boldsymbol{\theta}}{\arg \max } \sigma_{n}^{2} \\
		&=\underset{\boldsymbol{\theta}}{\arg \max } \frac{1}{m} \sum_{i} \sum_{k} \bm{W}_{i k}\left\|\boldsymbol{y}_{i}-\lambda_{f_{s_{k}}}\left(\boldsymbol{y}_{i}\right)\right\|_{2}^{2}
		\label{eq12}
	\end{aligned}
\end{equation} \\
The local optimal $\boldsymbol{\theta}$ can be obtained by the backpropagation algorithm. In general, E-step is to compute the soft decision information with foregone fading, and M-step is to estimate the channel with soft decision information.
\begin{figure*}[htbp]
	\centering
	\subfloat[]{
		\begin{minipage}[t]{0.15\linewidth}
			\centering
			\includegraphics[width=1in]{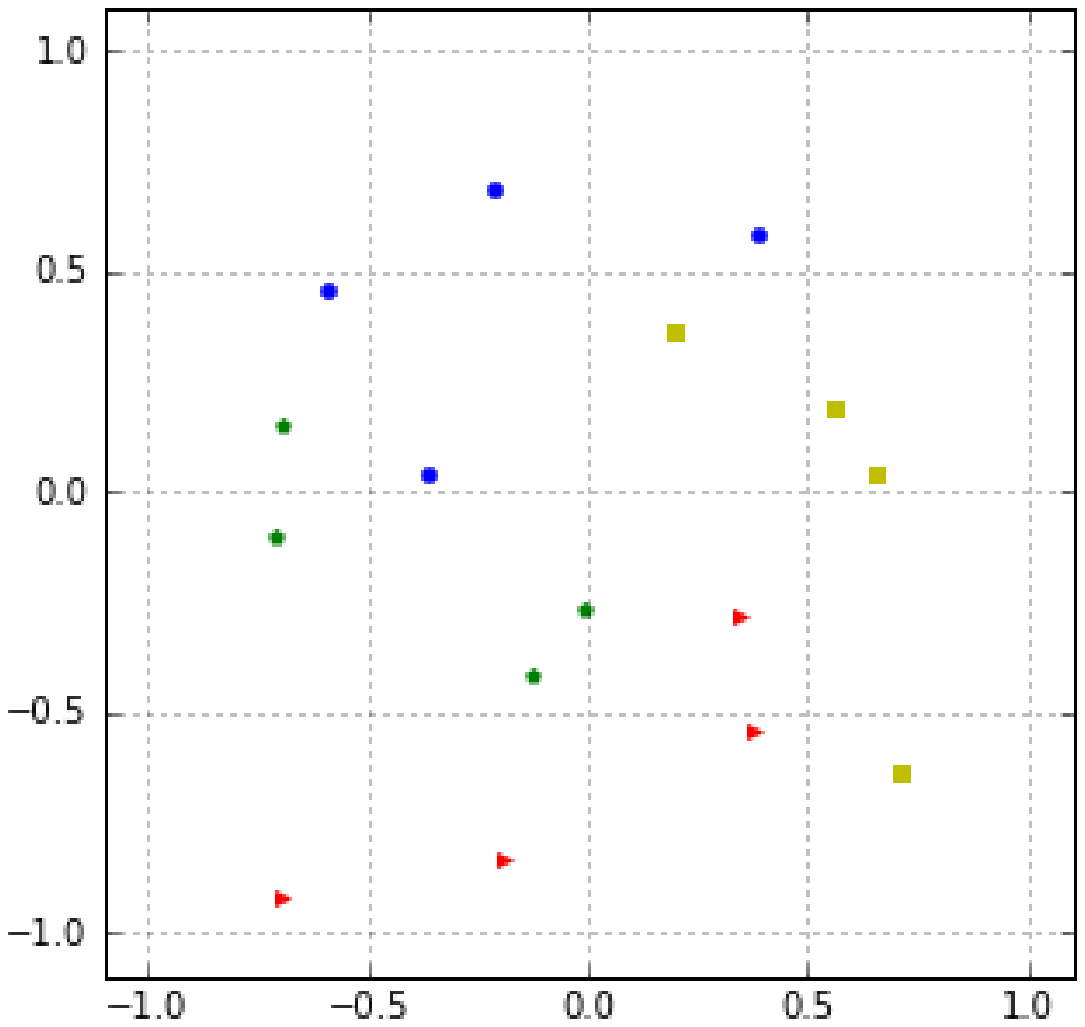}
			\label{fig3_a}
		\end{minipage}%
	}%
	\subfloat[]{
		\begin{minipage}[t]{0.15\linewidth}
			\centering
			\includegraphics[width=1in]{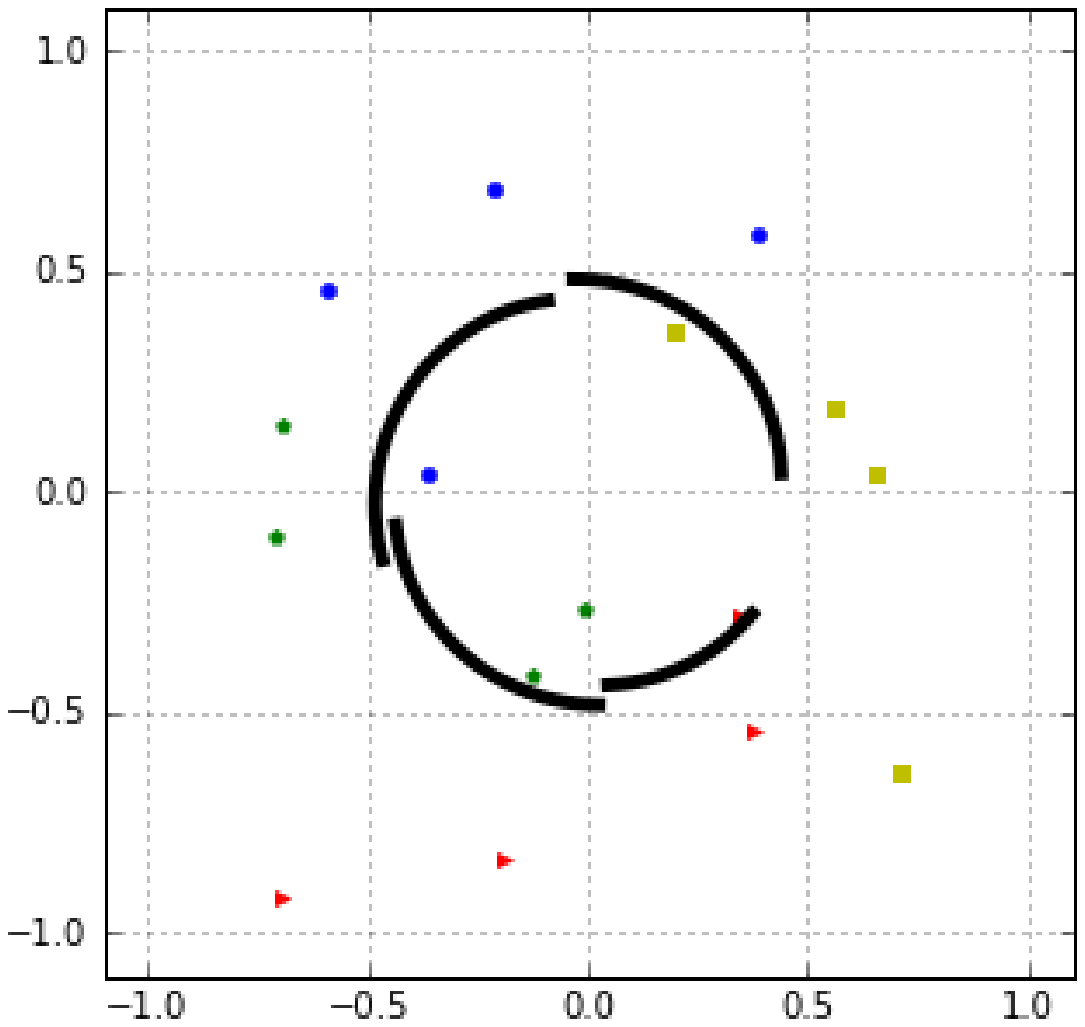}
			\label{fig3_b}
		\end{minipage}%
	}%
	\subfloat[]{
		\begin{minipage}[t]{0.15\linewidth}
			\centering
			\includegraphics[width=1in]{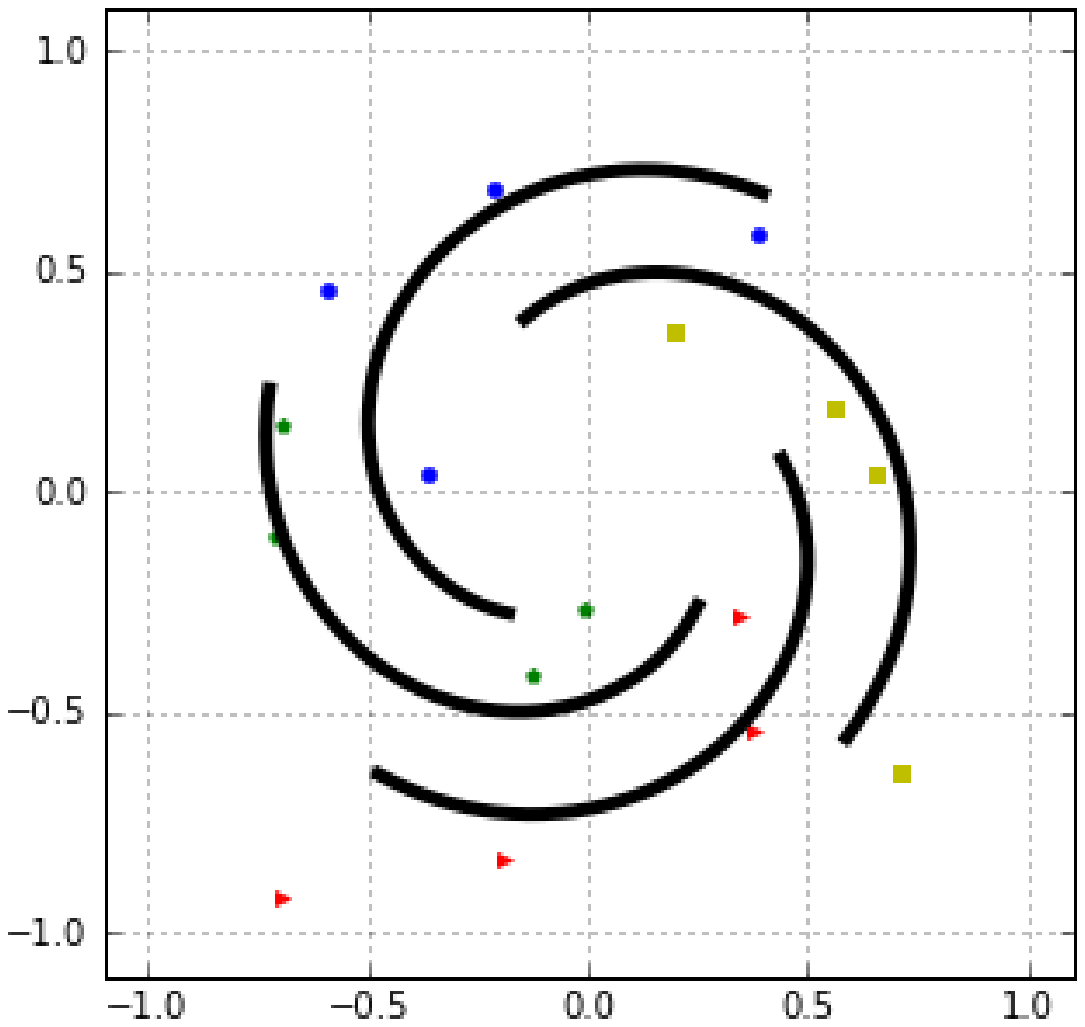}
			\label{fig3_c}
		\end{minipage}
	}%
	\subfloat[]{
		\begin{minipage}[t]{0.15\linewidth}
			\centering
			\includegraphics[width=1in]{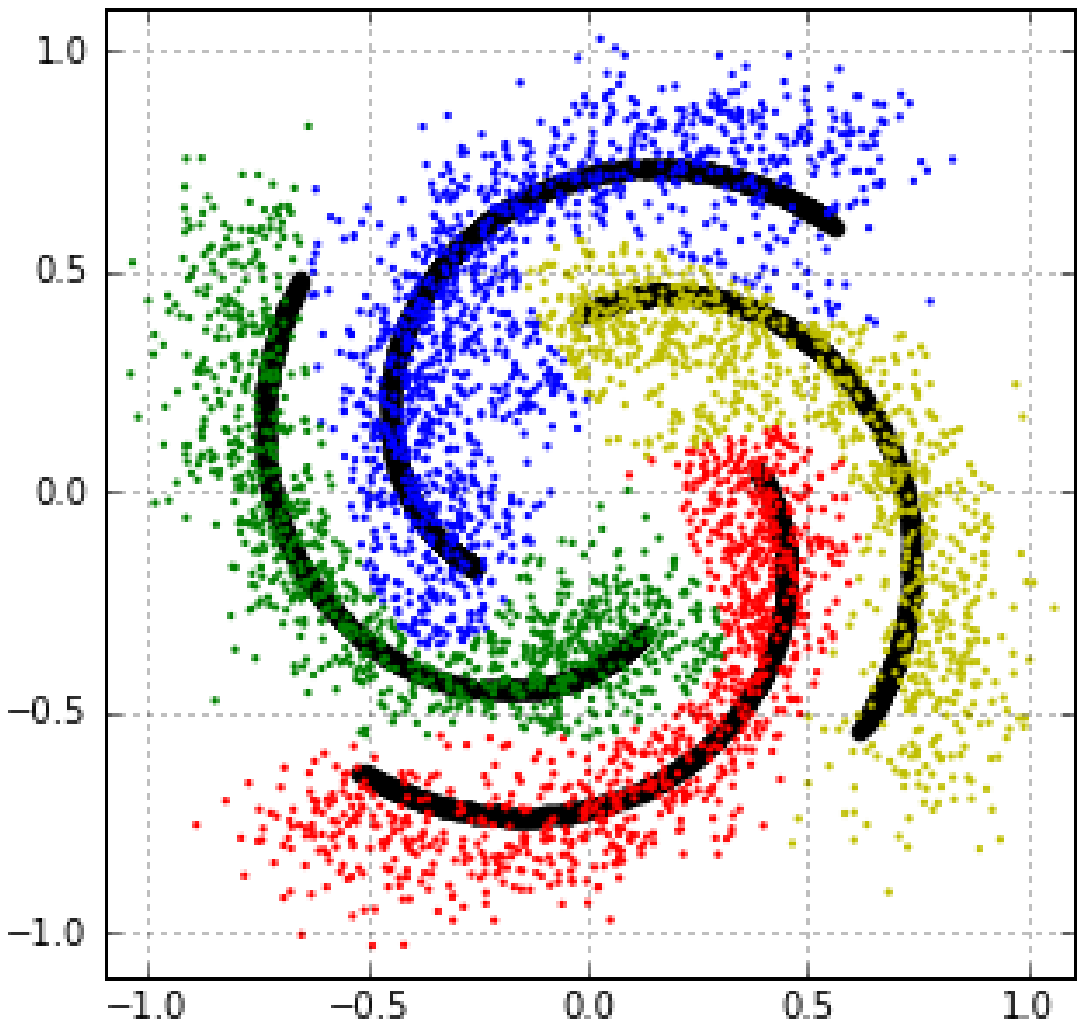}
			\label{fig3_d}
		\end{minipage}
	}%
	\subfloat[]{
		\begin{minipage}[t]{0.15\linewidth}
			\centering
			\includegraphics[width=1in]{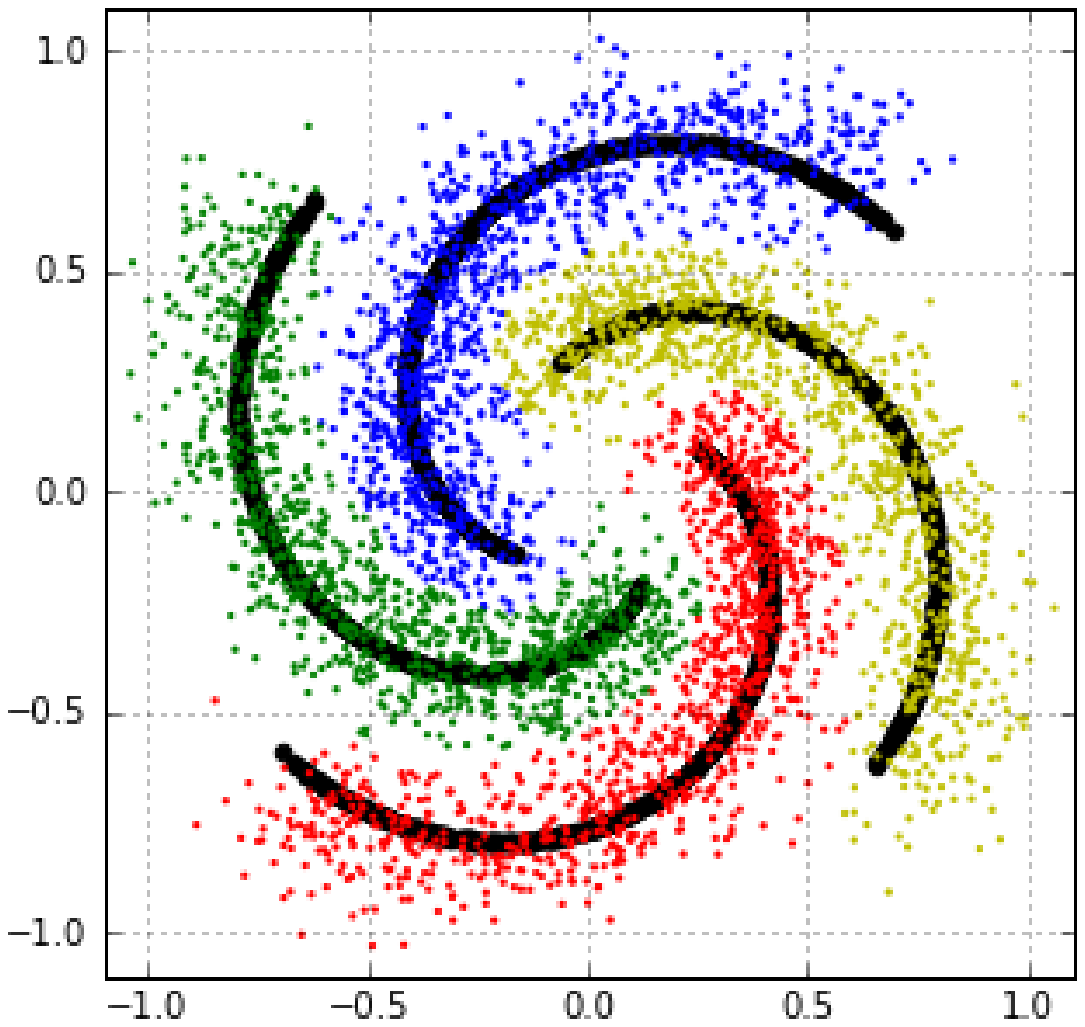}
			\label{fig3_e}
		\end{minipage}
	}
	\subfloat[]{
	\begin{minipage}[t]{0.15\linewidth}
		\centering
		\includegraphics[width=1in]{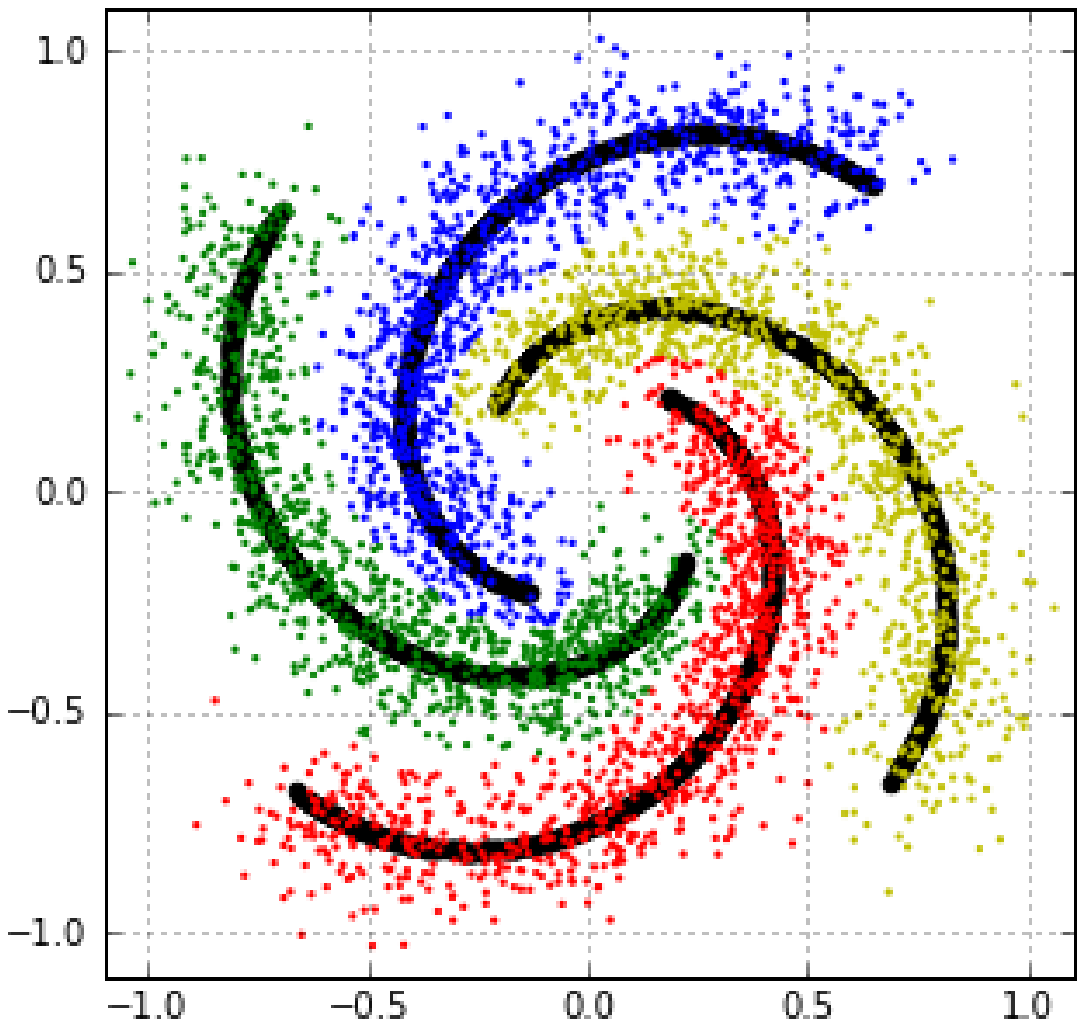}
		\label{fig3_f}
	\end{minipage}
	}%
	\centering
	\caption{The learning process of curve in network training.}
	\label{fig3}
\end{figure*}

\subsection{Symmetric Manifold Network}
\begin{figure}[!t]
	\centering
	\includegraphics[width=3.5in]{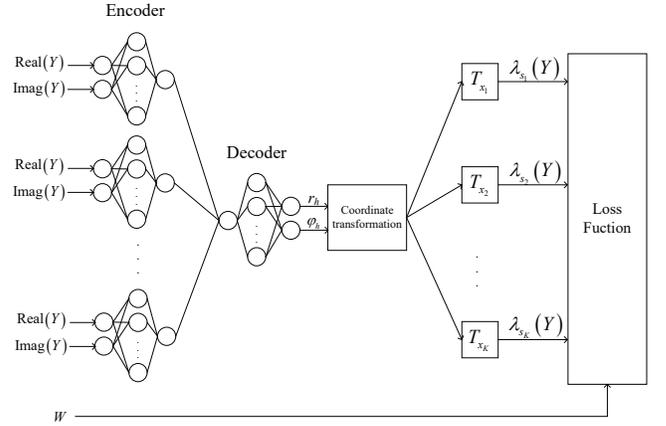}
	\caption{The structure of SMN}
	\label{fig2}
\end{figure}
Complex signal can be split into real and imaginary components, hence the input of SMN are two-dimensional real values. In our implement, The network first learns the polar coordinates of $\bm{h}$ and then converts it into Cartesian coordinates. See Fig.\ref{fig2}, each curve has a independent encoder and transformation matrix $T_{x_{i}}$ whereas decoder is the same. $T_{x_{i}}$ is given by
\begin{equation}
	T_{\bm{x}_{k}}=\rho_{\bm{x}_{k}} \left[ \begin{array}{cc}{\cos \varphi_{\bm{x}_{k}}} & {-\sin \varphi_{\bm{x}_{k}}} \\ {\sin \varphi_{\bm{x}_{k}}} & {\cos \varphi_{\bm{x}_{k}}}\end{array}\right]
	\label{eq13}
\end{equation}
where $\rho_{x_{k}}$ and $\varphi_{x_{k}}$ are respectively the magnitude and phase of $\bm{x}_k$. This trick guarantees the symmetry of the curves while reduced the requirement for the amount of training samples.

In summary, our SMN algorithm can be decomposed in following steps: Firstly, extract the uniformly inserted training sequence from the received signal. Then, employ the training sequence to pretrain the network with (\ref{eq12}). Finally, the entire sequence is trained using the EM algorithm with (\ref{eq10}) and (\ref{eq12}) iteratively. Refer to Section \RNum{4} for details of implementation.

\section{Experiment}
In this Section, we simulate the performance of SMN with the software environment that Python v3.6, Tensorflow v1.11.0 and Numpy v1.16.3. The channel parameters is set to carrier frequency $\omega=9$ GHz, collision frequency $v_{e n}=20$ GHz, and the electron density $n_{e} \in\left[1 \times 10^{16}, 6 \times 10^{17}\right]\mathrm{cm}^{-3}$. Since the ReLU and leakyReLU are not differentiable at 0, we use the tanh in the hidden layers. Both the middle layer of the encoder and decoder are 4 dimension. As for the output layer, in order to ensure that the amplitude of the fading is always less than 1, we use the sigmoid function to limit $\left|r_{h}\right| \leq 1$ and linear function for $\varphi_{h}$. The initializer of weights and biases is $\mathcal{N}\left(0, 0.1\right)$. The optimizer is Adam with an initial learning rate of $1 \times 10^{-3}$. The number of pretraining  steps, the iteration steps for the EM algorithm and  the training steps in each M-step is 2000, 10 and 100 respectively. Another trick we used is to initialize the optimizer before each M-step to reduce the probability of network convergence to a local optimal solution. The length of the transmitted sequence is 4096, in which the training sequence is evenly inserted at equal intervals.

\subsection{Learning Process}
In Fig.\ref{fig3}, with 256 insertion interval, we intercepted several states in  learning process of curve. The first figure shows the training sequence which does not show any distribution visually. The first two in the middle are in the pre-trained state, the next two are in the state of the EM algorithm iteration and the last one shows the final demodulation. It can be seen that the amplitude is almost fading to 0 and the phase is shifted by nearly 90 degrees. However, only 16 training samples are used to learn the general form of the fading curve in the pretraining. Whereafter the EM algorithm further approximates the curve and obtains perfect classification performance.

\subsection{SER Analysis}
\begin{figure}[!t]
	\centering
	\includegraphics[width=3.6in]{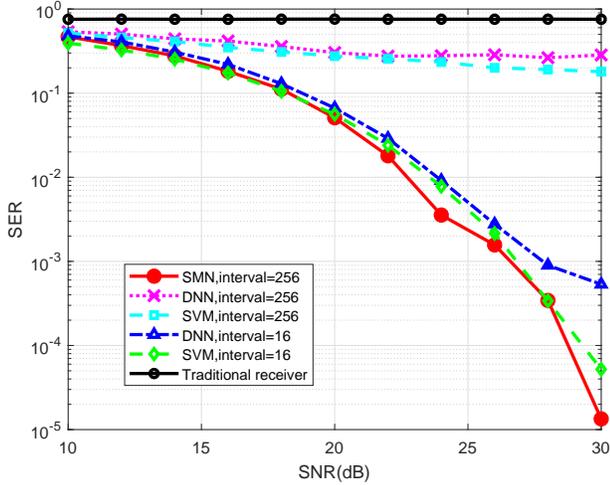}
	\caption{SER vs. SNR in QPSK system}
	\label{fig4}
\end{figure}

In Fig.\ref{fig4}, we compare the SER curves of various algorithms. Support vector machine (SVM) and DNN are the most advanced nonlinear models in supervised learning domain. Supervised learning is to find the optimal classification boundary in the label data. To achieve ideal generalization, sufficient samples are needed to describe the ground-truth distribution of the data. It can be seen that SMN obtains the desired performance at interval=256 (i.e., the bandwith utilization rate is 99.6$\%$). However, SVM and DNN maintain a high SER all the time. SVM and DNN have similar performance to SMN when interval=16 (i.e., the bandwith utilization rate is 93.75$\%$). That is, SMN dramatically reduces bandwidth consumption to 1/16. One of the significant reason is that semi-supervised learning could make full use of the unlabeled data in training. Another is that the symmetry trick is used to avoid learning each curve independently. 

\begin{figure}[h]
	\centering
	
	\subfloat[SNR=20]{
		\begin{minipage}[t]{0.3\linewidth}
			\centering
			\includegraphics[width=1in]{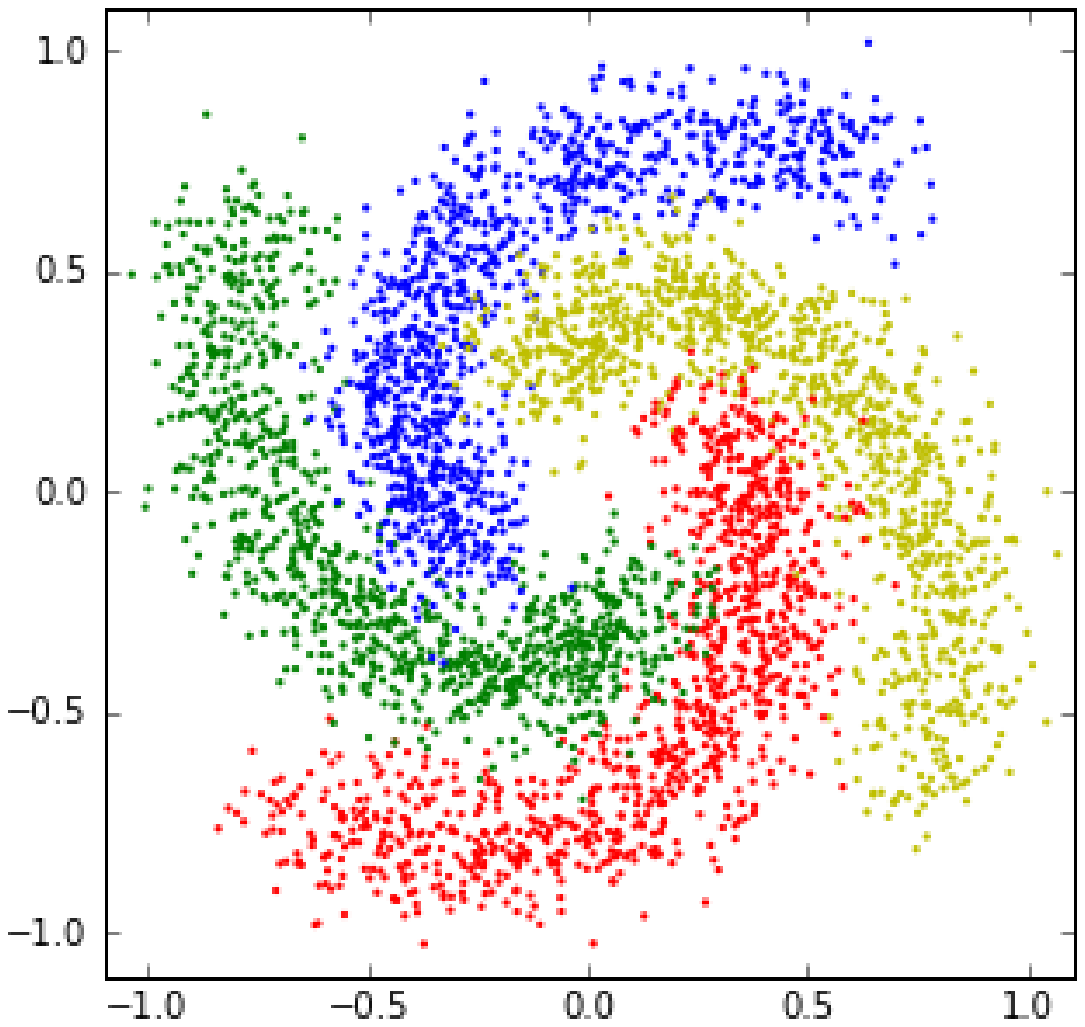}
		\end{minipage}%
	}%
	\subfloat[SNR=11]{
		\begin{minipage}[t]{0.3\linewidth}
			\centering
			\includegraphics[width=1in]{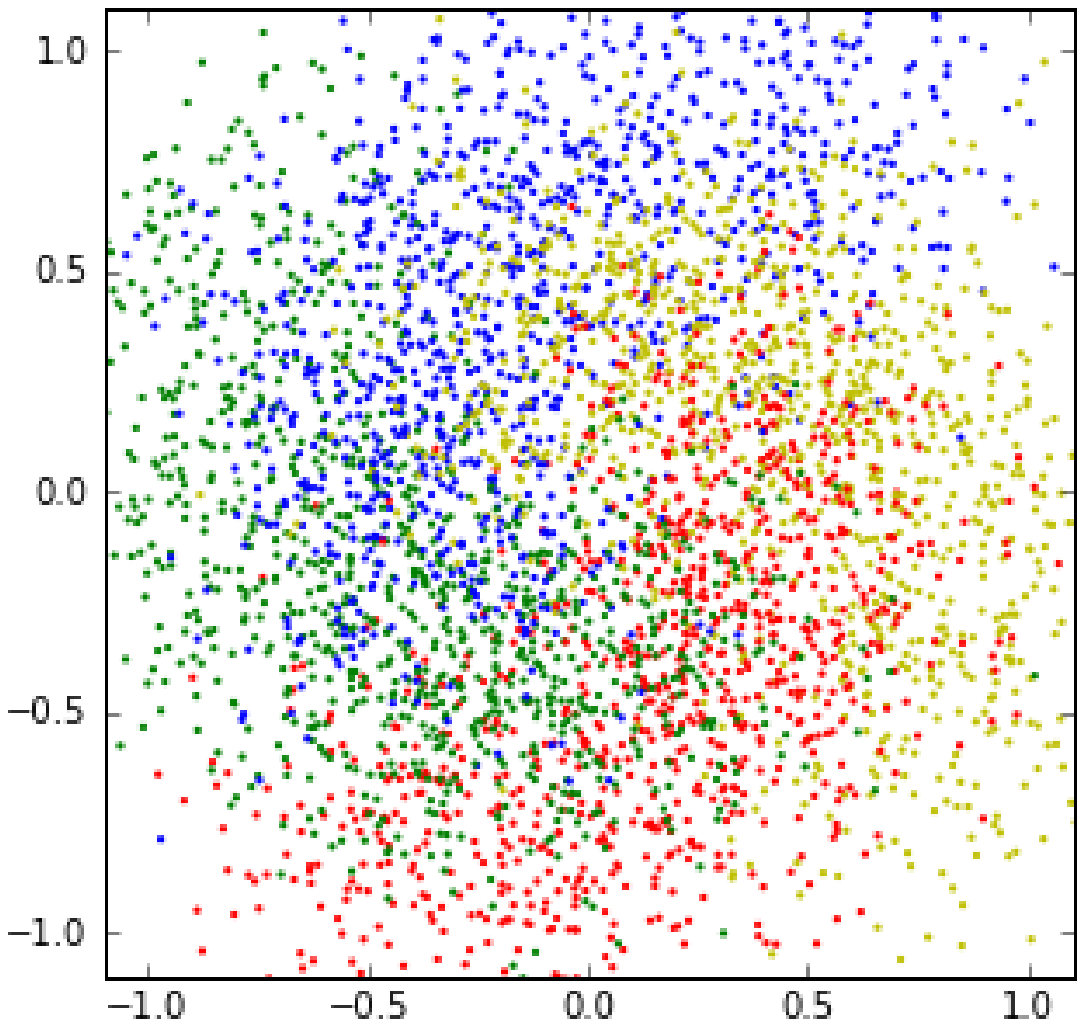}
		\end{minipage}%
	}%
	\subfloat[Result]{
		\begin{minipage}[t]{0.3\linewidth}
			\centering
			\includegraphics[width=1in]{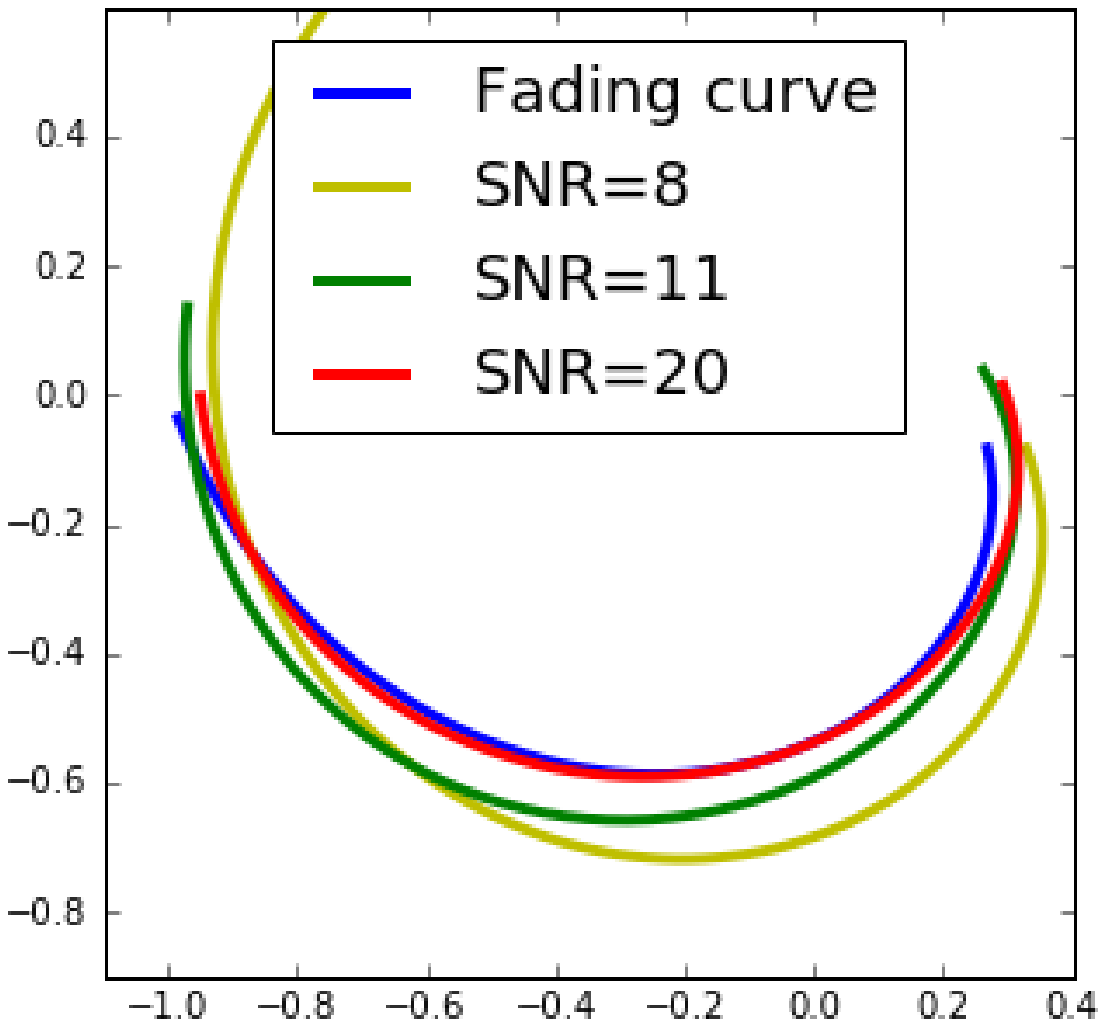}
		\end{minipage}
	}%
	
	\centering
	\caption{Fading estimation in various SNR}
	\label{fig5}
\end{figure}
\subsection{Fading Estimation}
Fig.\ref{fig5} shows the received signals which colored by ground-truth labels and the results of channel estimation in various SNR. It can be observed that the learned curve is almost identical to the fading curve when the SNR is high. There is a lot difference between the two ends of the curve because the data at both ends is sparse, and the SMN cannot capture the ground-truth distribution. While the SNR=11, although the received signals overlap, SMN can still perceive its weak law of distribution and the deviation of estimation is partly acceptable. As the SNR continues to decrease, the received signals are completely overlapping and the SMN cannot make a reasonable estimation.

\section{Conclusion}
In this paper, we propose a novel DL algorithm for the problem of amplitude fading and phase shift caused by plasma sheath. Joint optimization is achieved by extracting the principal components and coupling the estimation and demodulation. Simulation results show that while maintaining SER performance, we significantly reduce the bandwidth consumption. However, high performance leads to increased complexity. Although the training steps are few  compared with most of deep learning algorithms, the computational efficiency is not satisfied in communication system. Future research will concentrate on reducing the complexity of the algorithm and extending our approach in more scenarios.

\ifCLASSOPTIONcaptionsoff
\newpage
\fi

\bibliographystyle{IEEEtran}
\bibliography{cited}

\end{document}